\def\fun#1#2{\lower3.6pt\vbox{\baselineskip0pt\lineskip.9pt
  \ialign{$\mathsurround=0pt#1\hfil##\hfil$\crcr#2\crcr\sim\crcr}}}
\relax
\documentstyle[12pt]{article}
\advance\textheight by \topskip
\begin{document}
\begin{flushright}
SU-ITP-94-27\\
QMW-PH-94-34\\
hep-th/9409060
\end{flushright}

%\bf
\vspace{-0.2cm}
\begin{center}
{\large\bf EXACT  $SU(2)\times U(1)$  STRINGY BLACK HOLES}\\
\vskip 2 cm

{\bf Renata Kallosh \footnote { E-mail address:
kallosh@physics.stanford.edu} and
Tom\'as Ort\'\i n \footnote{ Permanent address: Department of
Physics, Queen Mary and Westfield College, \\
Mile End Road, London E1
4NS, U.K.
E-mail address: ortin@qmchep.cern.ch } }
 \vskip 0.1cm
Physics Department, Stanford University,
Stanford    CA 94305

\end{center}
\vskip 1.5 cm
\centerline{\bf ABSTRACT}
\begin{quotation}

Extreme magnetic  dilaton black holes are promoted to  exact solutions of
heterotic string theory with unbroken supersymmetry. With account taken of
$\alpha'$ corrections  this is accomplished by supplementing
the known solutions with  $SU(2)$ Yang-Mills vectors and scalars in addition to
the already existing
Abelian $U(1)$ vector field. The solution has a simple analytic form and
includes multi-black-holes.

The issue of exactness of other black-hole-type solutions, including extreme
dilaton electrically charged black holes and Taub-NUT solutions
is discussed.
\end{quotation}

\newpage

%%%%%%%%%%%%%%%%%%%%%%%%%%%%%%%%%%%%%%%
Charged four-dimensional extreme dilaton black holes \cite {GM} with
stringy
dilaton coupling $a=1$ have attracted a lot of
attention in recent years. In particular it was established that they have half
of  supersymmetries  unbroken when embedded into $d=4,\; N=4$ pure supergravity
without matter \cite{US}. As solutions of quantum four-dimensional
supergravity
they are subject to some supersymmetric non-renormalization theorem. In
particular, it has been argued that  the entropy of extreme black holes does
not get  quantum corrections \cite{US}, \cite{HULL}.

Besides considering these solutions in the framework of four-dimensional
quantum
supergravity one may address the issue of  stringy $\alpha'$ corrections for
these black holes.   The stringy $\alpha'$ corrections related to the Lorentz
anomaly in the general  context of
unbroken
supersymmetry of the heterotic string have been analyzed in \cite {Be1}.
For supersymmetric string
solitons and
multi-monopole solutions
 $\alpha'$ corrections  were studied
in  \cite{CHS}, \cite{Kh}. These solutions with $SU(2)$ Yang-Mills fields are
considered
to be exact solutions of the heterotic string theory.
The $\alpha'$ corrections of  generalized supersymmetric gravitational wave
solutions
as well as
dual wave solutions have been studied in \cite {BKO}, \cite {BEK}. These
solutions
require $SO(8)$ Yang-Mills fields for unbroken supersymmetry and exactness
in the heterotic string theory.
The extreme black
holes with
$a=\sqrt 3$  without Yang-Mills fields were found to be exact solutions of
the type II superstring theory in \cite{DUFF}.

The purpose of this letter is to investigate whether  $a=1$ charged dilaton
black
holes, which solve  the leading order equations of motion of the string
effective action, can be
promoted to exact solutions of   heterotic
string theory. We will try to  supplement them with a non-Abelian
Yang-Mills field that cancels the gravitational part of the
$\alpha'$ corrections. This will be an efficient use of the
 Green-Schwarz mechanism of the cancellation of  the Lorentz and Yang-Mills
anomalies, as well as supersymmetry anomaly \cite{K85}, \cite{Be1} for a given
configuration.
For this purpose it is useful to work directly
in the critical
dimension of  the heterotic string : $d=10$. The analysis of unbroken
supersymmetry as well as the issue of $\alpha'$ corrections is much simpler in
$d=10$. Therefore we will first uplift the four-dimensional dilaton
black holes to $d=10$ and analyze the $\alpha'$ corrections there. In the case
of the magnetic extreme dilaton black hole it will be possible to find the
appropriate
ten-dimensional Yang-Mills configuration to cancel the gravitational part of
the $\alpha'$ corrections. The non-Abelian field is completely  determined by
the
gravitational part of the uplifted black hole. By adding the Yang-Mills field
to the solution we will also achieve the restoration of unbroken supersymmetry
which
becomes anomalous \cite{K85}, \cite{Be1}  if only gravitational $\alpha'$
corrections of the heterotic
string are taken into account. Under plausible assumptions, the solution
obtained in this way is free  of $\alpha'$ corrections.

By the end of this analysis, we will dimensionally reduce our exact
solution to $d=4$. There we will get an additional $SU(2)$ scalar and vector as
part of the non-Abelian magnetic black hole.

 The issue of $\alpha'$ corrections for extreme magnetic dilaton black holes in
four-dimensional
geometry has
been studied
in \cite{Nat}, where it was shown that if the  gravitational
$\alpha'$ corrections are taken into account, the solution gets corrections. We
will see later that it is the same balance between the
gravitational $\alpha'$ corrections and Yang-Mills contribution responsible for
the cancellation of anomalies which has
allowed us to
preserve the
gravitational part of the black hole intact.

For extreme electric black holes  this procedure of promoting a lowest
order solution of a heterotic string theory to an exact one  fails. Our
procedure involves the embedding of the spin connection into a gauge group. The
corresponding gauge group for the electric dilaton black hole turns out to be
non-compact. The dilatonic  IWP solutions \cite{KKOT} could be made
exact if certain relations between the NUT charge and the mass
hold.

%%%%%%%%%%%%%%%%%%%%%%%%%%%%%%%%%%%%%%%%%%%

 Let us start with the extreme magnetic  dilaton black hole  \cite
{GM} in four dimensions. In this paper we will work  in stringy
frame\footnote{Our conventions are essentially those of Ref. \cite{BEK}, but
we reserve the indices $i,j,k,l,m$ for the three-dimensional space and
Yang-Mills indices. Yang Mills indices can be identified for being always the
last ones in upper position.  Also we underline curved indices and
$\epsilon_{123}=\epsilon^{123}=+1 $.}.
\begin{eqnarray}
ds^{2} & = & dt^{2}- e^{4\phi} d\vec{x}^{2}\, ,
\nonumber \\
F_{ \underline i  \underline k } &= & 2 \partial_{[ \underline i} V_{\underline
k ]} =  \pm  \epsilon _{ikl} \;\partial_{\underline l }
e^{2\phi} \, ,
\nonumber \\
e^{2\phi} & = & 1+ \sum_s \frac{2M_s}  {|\vec x - \vec x_s|}\, .
\label{magn} \end{eqnarray}
The Abelian vector field  $V  =  V_{\underline i}d x^{\underline i}$  is
defined up to a gauge transformation. For example, for a single black hole we
can take
\begin{eqnarray}
\vec V&=& \pm {2M ( 0, 0, z) \wedge  (x, y, z)  \over \rho (\rho^2 - z^2)}\,  ,
\qquad F_{ \underline i  \underline k } = \mp 2M \epsilon _{ikl} {x^{\underline
l} \over  \rho^3}
\, ,
\nonumber \\
e^{2\phi} & = & 1+ {2M \over \rho}\,   , \qquad  \rho^2 = \vec x^2 = x^2 + y^2
+ z^2 \, .
\end{eqnarray}
The ten-dimensional uplifted version of this geometry\footnote{The uplifting of
the single magnetic dilaton black hole in spherically symmetric coordinate
system was performed in \cite{NEL}.} can be easily obtained by
using  supersymmetric  dimensional reduction  \cite{Cham}, \cite{BKO2}.  In
the ten-dimensional configuration there are no vector fields, but there are
non-diagonal
components of the
metric as well as a two-form field.
\begin{eqnarray}
ds^{2}_{(10)} & = & dt^2 - e^{4\phi}d\vec{x}^{2}-(dx^{\underline 4} +
V_{\underline i} \, d x^{\underline i})^{2}-
dx^{\underline I}dx^{\underline I}\, ,
\nonumber \\
B_{(10)} &= &  {1\over 2} B_{\mu\nu} \; dx^\mu \wedge dx^\nu  = - V_
{\underline
i}\,  d x^{\underline i} \wedge dx^{\underline 4}
\, ,
\nonumber \\
\partial_ {\underline i }  \partial_{\underline i }  e^{2\phi} & = & 0\, ,
\qquad
e^{2\phi}  =  1+ \sum_s \frac{2M_s}  {|\vec x - \vec x_s|}\, .
\qquad
2 \partial_{[ \underline i} V_{\underline k ]} =  \pm  \epsilon _{ikl}
\;\partial_{\underline l }
e^{2\phi}
\label{upl}\end{eqnarray}
This uplifted extreme magnetic black hole only solves the zero-slope limit  of
the equations of
motion of the heterotic string effective action. Fortunately, as we will show
later, it is possible to promote this zero slope limit solution
to one exact  at all orders  in  $\alpha'$.  For this purpose the solution  has
to be supplemented
by
a ten-dimensional $SU(2)$ Yang-Mills field, which in this case is
\begin{equation}
A_{k}{}^ {i } = - \epsilon_{i kl} \partial_{\underline l }  e^{- 2\phi}  \qquad
A_{4}{}^ {i} = \pm 2  \partial_{\underline{i} }  e^{- 2\phi}\, .
\label{ym}
\end{equation}

In what follows we are going to prove that {\it the ten-dimensional
configuration (\ref{upl}),
(\ref{ym}) is an exact solution of the heterotic string theory effective action
with unbroken supersymmetry}.

%%%%%%%%%%%%%%%%%%%%%%%%%%%%%%%%%%%%%%%%%%%%%%%
The idea of identifying the gravitational spin connection with the
Yang-Mills
 field
was suggested long time ago \cite{W}.  It turns out to be extremely effective
in our
case.
In the context of  heterotic string theory the idea of the embedding of the
spin connection into the gauge group was developed in \cite{WCHS} and applied
 in \cite{CHS}, \cite{Kh}. We would like to stress however the important fact
discovered by Bergshoeff and Roo \cite{Be1}  that there are two spin
connections with torsion, $\Omega_{\pm} =\omega(e) \mp  \frac{3}{2}H$ that play
different roles: $\Omega_{+}$ controls the classical part of
supersymmetry
transformation
rule of gravitino, whereas
 $\Omega_{ - }$ is responsible for
$\alpha'$ corrections to the action and to the supersymmetry rules,
related to Lorentz and supersymmetry anomalies. Because of that,  the proper
embedding of the spin connection into the gauge
group is
\begin{equation}
\Omega_{\mu - }{}^{ab} =  A_\mu{} ^{ab}\ .
\end{equation}
This means that it is $ \Omega_{\mu - }{}^{ab}$ which  has to be identified
with the gauge field to cancel  anomalies as well as corrections to the
supersymmetry rules and to solutions.
 Heterotic string $\alpha'$ corrections to supersymmetry
transformation
laws  as well as to equations of motion enter via $T$-tensors and
Chern-Simons
terms, defined in \cite{Be1} and described in detail in \cite{BKO}, \cite{BEK}.
If the corresponding
$T$-tensors and Chern-Simons terms  defined in eqs. (49)-(52)
of
\cite{BKO}  do not
vanish for a given configuration, the
 supersymmetry of a given configuration which is unbroken in the lowest order
 is destroyed  unless the torsionful spin connection
$ \Omega_{\mu - }{}^{ab}$ is
embedded into the gauge group.  In our previous work
\cite{BKO}
we have shown that {\it there are two necessary conditions
for having
unbroken supersymmetry and absence of  $\alpha'$ corrections to a
given configuration}:

i) The configuration has to solve the lowest-order equation of motion. The
corresponding Yang-Mills equations  (\ref{YM}) will be defined below.

ii) All $T$-tensors and Chern-Simons terms have to vanish for the
configuration.

For the special case  of  gravitational waves, supplemented by a two-form and
non-Abelian field, which we called SSW \cite{BKO},  we have shown that both
conditions
above
are met. We have also found that for the most general of those solutions  the
Yang-Mills
group is $SO(8)$.

Here we  start with the configuration without Yang-Mills field which
solves the lowest-order equations of motion for the metric, dilaton and the
two-form field.
The lowest-order equation of motion for the
Yang-Mills
field
follows from the effective action of the heterotic string which
includes first
order
$\alpha'$ corrections.
\begin{equation}
S^{(1)} =\frac{1}{2}\int d^{10}x
e^{-2 \phi}\sqrt{- g} \, [ -R+
4 (\partial\phi )^{2}-\frac{3}{4}H^{2} + \frac{1}{2} T]\, ,
\label{eq:actionD2}
\end{equation}
where
\begin{equation}
T =  2  \alpha^{\prime} \, \biggl (
R_{\mu\lambda}{}^{ab}\bigl
                         (\Omega_-)
                R^{\mu \lambda } {}_{  ba}(\Omega_-)
                -  \, {\rm tr} F_{\mu  \lambda}
                F^{\mu \lambda} \biggr )\label{eq:t2}\ .
\end{equation}
The trace is in the  vector representation of the $SO(32)$ group
 and the $H^2$-term does contain the first order correction term in
$\alpha^{\prime}$ given by the combination of Yang-Mills and Lorentz
Chern-Simons terms. The detailed form of the
Yang-Mills equation is given in eq.
(62) of  \cite{BKO}, where also other details of the procedure,
described
here shortly, can be found.  The lowest-order
Yang-Mills equation can be presented in a very compact form
\begin{equation}
\alpha'  \; D_a{}^+ (e^{-2 \phi} F^{ab}) = 0 \, ,
\label{YM}\end{equation}
where $D_a{}^+$ is the Yang-Mills and general covariant derivative
associated with the  connections  $A_\mu$ and $\Omega_{\mu
+}{}^{ab}$.

 In what
follows, we are
going to exhibit the corresponding spin connections for the uplifted extreme
magnetic dilaton
black hole and use them to rescue the unbroken supersymmetry, which is
damaged by
the gravitational part of $\alpha'$ corrections not balanced by the
Yang-Mills
field contribution.

Our configuration corresponds to a decomposition of the manifold
$M^{1,9}
\rightarrow M^{1,5} \times M^4$ and the tangent space
$SO(1,9)\rightarrow
SO(1,5) \times SO(4)$.  The curved manifold is only $M^{4}$ since $
M^{1,5}$ is
flat. The  zehnbein basis is given by the  sixbein part  $e_{\underline t}{}^t
= e_{\underline I}{}^I = 1$  and the vierbein part    $e_{\underline i}
{}^j = e^{
2\phi}\delta _{\underline i}{} ^j , \;
e_{\underline i}{} ^4= V_{\underline i}, \; e_{\underline 4}{}^i = 0,
\;
e_{\underline 4}{}^4=1$.
The non-vanishing components of  $\Omega_{c \pm }{}^{ab}$ in the above
 basis are
\begin{equation}
\Omega_{4 - }{}^{ij} = \Omega_{i + }{}^{4j}  = \pm \epsilon _{ijk}
\partial
_{\underline k}
e^{-2\phi}  \, , \qquad            \Omega_{k - }{}^{im} = \Omega_{k +
}{}^{im}=
2
\delta_{k[i}   \partial _ {\underline m]} e^{-2\phi} \, .
\label{SC}\end{equation}
Having found the torsionful spin connection we may perform the
embedding of the
spin connection into the gauge group. We will relate the spin
connection to a
gauge field $A_\mu ^k$
as follows
\begin{equation}
\Omega_{\mu - }{}^{ij} = \epsilon^{ijk} A_\mu ^k \ .
\label{embed}\end{equation}
The choice of the gauge group is $SU(2)$, according to eqs.
(\ref{SC}). This
gauge group is a  subgroup  of the gauge
group of the heterotic string $SO(32)$ (or $E_8\times E_8$).

Using the spin connections defined above we have confirmed the
unbroken
supersymmetry in the zero slope limit for the uplifted black holes
directly in
terms of the ten-dimensional gravitino and dilatino supersymmetry
rules. The
Killing spinor is a constant spinor, satisfying one of the
constraints
($\gamma_5 \equiv \gamma_1 \gamma_2 \gamma_3 \gamma_4$)
\begin{equation}
(1 \pm \gamma_5) \epsilon _{\pm} = 0 \, .
\label{constr}\end{equation}
The integrability condition for the unbroken supersymmetry is
expressed in the
simplest way as the  (anti)self-duality condition for the torsionful
curvatures
in  $M^4$.
\begin{equation}
R_{\mu\nu}{}^{ij} (\Omega_+) = \mp \epsilon_{ijm} R_{\mu\nu}{}^{m4} (\Omega_+)
\, .
\end{equation}
By performing the embedding we  complied with the
condition
ii). However,
how do we know that the spin-connection of the uplifted magnetic
black hole
does solve the Yang-Mills equation of motion? In view of the fact
that
non-Abelian black hole solutions have not  yet been constructed  in an
analytic
form, we  have to study this problem very carefully.
 Callan, Harvey and
Strominger  \cite{CHS} have made a crucial observation about the embedding
spin connection  into the gauge group. Due to the exchange
identity for
the torsionful curvatures, the
embedding automatically guarantees that the unbroken supersymmetry of the
solution remains unbroken after the addition of  the
Yang-Mills
field. Indeed by identifying $\Omega_{c - }{}^{ab}$ with the
Yang-Mills field
we checked directly that the gaugino supersymmetry rule also has an
unbroken
supersymmetry with the Killing Spinor satisfying   constraint  (\ref{constr}).
The
(anti)self-duality of the non-Abelian
field strength
\begin{equation}
F_{ik}{}^{j} = \mp \epsilon_{ikm} F_{m4}{}^{j}
\end{equation}
is also valid ( it follows from $  R_{ij}{}^{\mu\nu} (\Omega_-) = \mp
\epsilon_{ijm} R_{m4}{}^ {\mu\nu} (\Omega_-)
$)  and provides the integrability condition for the
unbroken
supersymmetry of gaugino.  However, supersymmetry in general may be
not
sufficient   to show that the configuration above
solves the
Yang-Mills equation of motion. We have checked that for our configuration
unbroken supersymmetry
does mean that the Yang-Mills equation of motion is satisfied.  We have also
checked directly that our solution (\ref{upl}), (\ref{ym}) solves eq.
(\ref{YM}) \footnote{The correct normalization for the $SU(2)$
      Lie algebra with our conventions is
      $[T_{i},T_{j}]=-\frac{1}{2}\epsilon_{ijk}T_{k}$.}.
The lowest-order equations of
motion for
the metric, dilaton
and two-form field
are satisfied, since we have performed a supersymmetric uplifting of
the
magnetic black hole.
Therefore
the condition  i)  of exactness is met for our solution.

Thus we  conclude that under the assumptions specified in \cite {BKO},
\cite {BEK}
our ten-dimensional configuration (\ref{upl}), (\ref{ym}), besides
being a solution
of the lowest-order equations of motion,  is an exact solution with unbroken
supersymmetry of the
effective
equations of motion of the heterotic string to all orders in
$\alpha^{\prime}$. The main assumption about the exactness of a solution is the
following.  A
solution which has unbroken supersymmetry is subject to
a supersymmetric
non-renormalization theorem in absence of supersymmetry anomalies.
The known
supersymmetry anomaly \cite{K85}, related to  Lorentz anomaly, is taken
care of by
 embedding the spin connection in the gauge group,  which results in the
necessary presence of a
specific
non-Abelian field in the solution.

 To support our space-time arguments  for the exactness of this solution we
have analysed  the appropriate superconformal world-sheet sigma model
following the analysis performed in  \cite{CHS}. By using the work of Howe and
Papadopolus \cite{HOWE} we have found that the corresponding sigma model  for
our special background has $(4,1)$ world-sheet supersymmetry.
The model can be presented in terms of  $(1,1)$ superfields \cite{HOWE}. For
this model the additional 3 supersymmetries exist  when the background
satisfies some conditions including the existence of 3 complex structures. The
complex structures are constructed from the chiral Killing spinors, responsible
for the unbroken space-time supersymmetry of our background. The bilinear
combinations of these spinors are covariantly constant with respect to the
proper
spin connection. The self-duality of the gauge field strength, which is also a
condition for additional world-sheet supersymmetry of our class of backgrounds,
is also met. Thus the sigma model discussed above for our special background
has
$(4,1)$ world-sheet supersymmetry. In the  heterotic string theory, this
property
of the background is believed to provide the exactness property of
the background \cite{HOWE}, \cite{CHS}.

%%%%%%%%%%%%%%%%%%%%%%%%%%%%%%%%%%%%%%%%%%%

The analogous procedure can be performed for other black-hole-type
solutions, known to be supersymmetric in  $d=4$. The most surprising
result of this analysis  concerns the extreme electric black hole. The extreme
electric dilaton black hole was uplifted to $d=10$ in \cite{BKO2}.
Meanwhile, Horowitz and Tseytlin have proved that the extreme electric dilaton
black
holes as well as their generalizations \cite{KKOT} are  exact solutions of
bosonic string theory \cite{HT}.  In
heterotic string theory we know that after we get the torsionful spin
connection of the uplifted configuration we have to identify the corresponding
non-Abelian field to nullify the above mentioned $T$-tensors and the
combination of Lorentz and Yang-Mills Chern-Simons terms to restore the
unbroken supersymmetry.  The corresponding non-vanishing components of the spin
connection
turn out to be
\begin{equation}
\Omega_{4 - }{}^{0i} = \pm  e^{4\phi} \partial _i  e^{-2\phi}  \, .
\label{SCel}\end{equation}
The Yang-Mills field which will serve to restore the unbroken supersymmetry and
fix
$T_{44} = T =0$ has to belong to a boosts  part of the non-compact group
$SO(1,3)$. This is not
part of
$SO(32)$ or $E_8\times E_8$ gauge system of the heterotic string in critical
dimension. Alternatively, the non-Abelian gauge field may belong to  $SU(2)$
group, but the field has to be imaginary. Both solutions do not seem to be
acceptable.
The black hole-wave  duality analysis  performed in \cite{BKO2} had  the
strange
feature that the electric black hole was dual to a wave with an imaginary
component in the metric. This mysterious property has found a complete
explanation now. We have found the system of coordinates where the
corresponding wave solution,
promoted to the exact one in the presence of the gauge fields, had the metric
and the 2-form field real, but the gauge field is still  imaginary.
This we  have found now from the direct analysis without using black hole-wave
duality. In conclusion,  the uplifted electrically charged black hole can not
be promoted to an exact solution of the heterotic string theory via spin
embedding, does not have an unbroken supersymmetry
and does not solve the field equations with account taken of the
$\alpha'$ corrections  in the frame for which the  the space-time
supersymmetric embedding is known \cite{Be1}. The situation seems to be better
for  the uplifted IWP axion-dilaton geometries \cite{KKOT},  at least for NUT
charges which are not smaller than the mass of the solution.  Our conclusion
about electric black holes is not in disagreement with the  analysis of
Horowitz and Tseytlin \cite{HT}. Indeed, we were looking for
exact solutions of heterotic string theory with unbroken supersymmetry which
are supposed to be stable.
 They have found that by changing the
renormalization scheme for the $\beta$-functions one can find a possibility
 without a gauge field to
keep the zero slope limit of these solutions exact due to
the null properties of one of the spin connections. The issue of
$\alpha'$ corrections to unbroken supersymmetry of the solutions
without a gauge field remains unclear, since it is not known if the effective
action which was  used in  \cite{HT} has a supersymmetric embedding. We are
trying to understand the relation between those two effective actions.

%%%%%%%%%%%%%%%%%%%%%%%%%%%%%%%%%%%%%%%%%%%%%%%
As we have seen, in ten dimensions the simple addition of an $SU(2)$
Yang-Mills field is enough to turn the uplifted extreme magnetic dilaton
black hole into an exact solution. The resulting configuration will also
be exact upon reduction to four dimensions. The four dimensional metric,
axion, dilaton and $U(1)$ vector field are unchanged. They are still
those of the leading order solution. But, in addition to them, the four
dimensional exact solution has an $SU(2)$ scalar $\Phi^{i}$  whose origin is
the fourth component of the ten-dimensional gauge field
$A_{\underline{4}}{}^{i}$, and an
$SU(2)$ vector field that we call $W$ to distinguish it from
the ten-dimensional $A$ whose $1,2,3$ components are identical in flat
indices but not in curved. This is due to the presence of non-diagonal
elements $V_{\underline{i}}$ in the ten-dimensional metric (\ref{upl}).

The four-dimensional exact $SU(2)\times U(1)$ dilaton (multi)-black
hole is, therefore
\begin{eqnarray}
ds^{2} & = & dt^{2} -e^{4\phi}d\vec{x}^{2}\, ,
\nonumber \\
e^{2\phi} & = & 1 +\sum_{s}\frac{2 M_{s}}{|\vec{x}-\vec{x}_{s}|}\, ,
\nonumber \\
F_{\underline{i} \underline{k}}(V) & = &  \pm\epsilon_{ikl}
                                          \partial_{\underline{l}}
                                          e^{2\phi}\, ,
\nonumber \\
\Phi^i & = & \pm 2 \partial_{\underline{i}}e^{-2\phi}  \, ,
\nonumber \\
W_{j}{}^i  & = & -2 \epsilon_{ijk}\partial_{\underline{k}}
                   e^{-2\phi} \, .
\end{eqnarray}

Since the dimensional reduction has been done in a way compatible with
supersymmetry, the four-dimensional exact $SU(2)\times U(1)$ black hole will
have
half of unbroken supersymmetries of $N=4$ supergravity interacting with $N=4$
Yang-Mills theory with account taken of stringy $\alpha'$ corrections.

As different from the pure $SU(2)$ solutions presented in Ref.
\cite{Kh}, the self-duality of the $SU(2)$ gauge field
strength in the Euclidean four-dimensional space $x^{1},\ldots,x^{4}$
does not translate into the conventional form of the Bogomolnyi bound
but into an analogous balance of forces expression that involves the
$U(1)$ field strength $F$:
\begin{equation}
G_{jk}{} (W)+\Phi  F_{jk}(V) = \pm\epsilon_{jkl}D_{l}\Phi \, .
\end{equation}
Again this is due to the non-diagonal elements of the ten-dimensional
metric (the $U(1)$ vector field $V$) which do not occur in the  ansatzes
used in Ref. \cite{Kh}.

The fields of a single $SU(2)\times U(1)$ black hole  are
\begin{eqnarray}
ds^{2} & = & dt^{2}- e^{4\phi} d\vec{x}^{2}\, ,
\nonumber \\
F_{\underline{i} \underline{k}} & = & P \epsilon_{ikl} \;
                                \frac{x^{\underline{l}}}{\rho^{3}}\, ,
                                \qquad P=\mp 2M\, ,
\nonumber \\
e^{2\phi} & = & 1+\frac{2M}{\rho}\, ,
\nonumber \\
\Phi^i & = & -2P \frac{x^{\underline{i} } }{\rho (\rho + 2M)^{2}}\, ,
\nonumber \\
W_{j}{}^i & = &  \mp 2 P \epsilon_{ijk}   \frac{x^{k} }{\rho (\rho +
2M)^{2}} \ .
\label{exmagn}
\end{eqnarray}
The asymptotic behavior of our non-Abelian fields is the same as  in stringy
monopole solution \cite{Kh}, they fall off faster than the fields of
't Hooft-Polyakov monopole. By performing a gauge transformation of the type
$g\partial _{\underline 4}
g^{-1}$ one may also change the asymptotic value of the scalar filed so that it
does
not vanish at $\rho\rightarrow \infty$.

The conclusion is that  by addition of an appropriate Yang-Mills field  in some
cases it is possible to get  solutions of the heterotic string effective action
 with unbroken supersymmetry  which are exact at all orders in
$\alpha^{\prime}$.  This we  have done   for the
extreme magnetic dilaton black holes and we have found  that the same procedure
does not work for the electric ones and for some of Taub-NUT solutions. The
procedure which we have
used can be applied  to other four-dimensional solutions
with unbroken supersymmetry. Our expectation is that all of these solutions
which will be found exact can be obtained in a closed analytic form. One may
hope that in this way more
information will be obtained about axion-dilaton black hole-type solutions and
new
restrictions on their parameters
may emerge. This may lead to better understanding of the charge quantization of
the black holes due to the possibility that the non-Abelian charges will play
an important role.

We are grateful to E. Bergshoeff,  G. W. Gibbons and A. Linde for extremely
stimulating discussions. We are grateful to the referee for the suggestion to
consider the world-sheet supersymmetry of the sigma model, corresponding to our
non-Abelian black hole.
This  work was  supported by  NSF grant PHY-8612280 and the work of T.
O. was supported also by  European Communities Human Capital and Mobility
programme
grant. T.O. would like to express his gratitude to the Physics Department of
Stanford University for its hospitality.

\end{document}